\documentclass[12pt]{article}

\usepackage{epsfig}
\usepackage{xcolor}
\usepackage{indentfirst}

\usepackage{amsmath}
\usepackage{bm}
\usepackage{natbib}
\usepackage{threeparttable}

\title{\bf
Efficient Computation of the Stochastic Behavior of  Partial Sum Processes}
\author{
Sorawit Saengkyongam\\
Department of Statistics\\
Chulalongkorn University\\
Bangkok, Thailand\\
\\
Anthony  Hayter\\
Department of Business Information and Analytics\\
University of Denver\\
Denver, USA\\ 
\\
Seksan Kiatsupaibul\\
Department of Statistics\\
Chulalongkorn University\\
Bangkok, Thailand\\
\\
Wei Liu\\
S3RI School of Mathematics\\
University of Southampton\\
Southampton, UK\\
 }

\date{ }

\begin{document}

\maketitle


\vspace{6mm}

\renewcommand{\abstractname}{Abstract}
\begin{abstract}
  \normalsize
In this paper the computational aspects of probability calculations for dynamical partial sum expressions are discussed.
Such dynamical partial sum expressions have many important applications, and examples are provided in the fields of reliability, product quality assessment, and stochastic control.
While these probability calculations are ostensibly of a high dimension, and consequently intractable in general, it is shown how a recursive integration methodology can be implemented to obtain exact calculations as a series of two-dimensional calculations.
The computational aspects of the implementaion of this methodology, with the adoption of Fast Fourier Transforms, are discussed.

\end{abstract}

\noindent {\it Key words}: Independent variables; partial sums; recursive computations; cumulative sums; conditional probability;
moments; reliability; product quality assessment;  stochastic control; Fast Fourier Transform.

\section{Introduction}\label{sec:intro}

The tracking of the stochastic behavior of a partial sum process is an important problem with many applications.
In general, calculations of the probabilistic properties of such a partial sum process require an ability to compute high-dimensional multivariate probabilities of 
partial sum variables.  Computing these multivariate probabilities is in fact a high-dimensional integration problem 
which in general cannot be performed efficiently by any numerical method presently available.  

Computing the multivariate probability of an event in high dimensions, 
in its most general form, is an intractable problem.  
An efficient solution may possibly be devised by exploiting any special structures of the problem under 
consideration.
For example, with a general high dimensional density function, 
if the event is a convex set, a Markov chain Monte Carlo approach can be devised to efficiently approximate the probability of the event (Smith (1984), Belisle et al. (1993), Lov{\'a}sz (1999) 
and Kiatsupaibul et.al. (2011)).  
On the other hand, if the probability distribution is a multivariate standard normal or $t$-distribution with 
some special correlation structure, an efficient numerical integration may be constructed to 
compute a rectangular event, say (Dunnett \& Sobel (1955), Soong \& Hsu (1997)), 
or an event based on a complete ordering (Kiatsupaibul et al.(2017)).

The partial sums of independent 
random variables have a nice structure 
that can be exploited to devise an efficient numerical algorithm for the calculation of their probabilistic properties.  The objective of this paper is to illustrate how such probability calculations for the stochastic behavior of a partial sum process of independent variables can be performed efficiently based upon the adoption of recursive numerical integration techniques.

The specific problem considered in this paper can be described as follows.
Let $X_i$, $1 \leq i \leq n$ be independent random variables.
In general, we consider probabilities of the form
\[
P((X_1, \ldots ,X_n) \in A ) =
\]
\begin{eqnarray}
P( (X_1,X_2,X_3) \in A_1, (X_1+X_2,X_3,X_4) \in A_2, \ldots , (X_1+ \ldots + X_{n-2},X_{n-1},X_n) \in A_{n-2} )
 \label{1}
\end{eqnarray}
for sets $A_i \subseteq \Re^3$, $1 \leq i \leq n-2$.
The methodologies discussed in this paper are applicable to the evaulation of this general expression.

However, a special and important case of equation (\ref{1}) is the sole consideration of the sum of the random variables
\begin{eqnarray}
P( X_1+ \ldots + X_{n} \in B )
 \label{2}
\end{eqnarray}
for a set $B \in \Re$ which has many varied applications.
When the sum of the random variables does not have an identifiable distribution, the evaluation of this probability is ostensibly challenging,
although it is shown in this paper how its evaluation is in fact straightforward for any value of $n$.

More generally, probabilities concerning the stochastic behavior of the partial sum process of the random variables of the form
\begin{eqnarray}
P( X_1 \in B_1, X_1+ X_{2} \in B_2, \ldots ,  X_1+ \ldots + X_{n} \in B_n )
 \label{3}
\end{eqnarray}
for sets $B_i \subseteq \Re$, $1 \leq i \leq n$, are also a special case of equation (\ref{1}).
In this paper it is also shown how the evaluation of this expression is in fact also straightforward for any value of $n$.

The key result of this paper is that the $n$-dimensional integral expression
\begin{equation}\label{4}
\idotsint\limits_{(x_1,\ldots,x_n) \in S} h_1(x_1)\cdots h_n(x_n)~dx_1\cdots dx_n
\end{equation}
can be evaluated recursively as a series of $2$-dimensional integral calculations when the set $S \subseteq \Re^n$
is defined by the conditions
\[
(x_1+ \ldots + x_i, x_{i+1},x_{i+2}) \in I_{i} \subseteq \Re^3
\]
for $1 \leq i \leq n-2$.
This is an application of the general discussion of recursive integration given in Hayter~(2006) with $d=2$.
Recursive computational techniques similar to the ones developed in this paper have been applied to the problem of confidence band construction for a distribution function in Kiatsupaibul \& Hayter~(2015), and to ranked constrained computations in Kiatsupaibul et al.~(2017).

Of course, the probability in equation 
(\ref{1}) can be put in this form for continuous random variables
with the sets $I_i$ equal to the sets $A_i$ and the functions $h_i(x_i)$ equal to the 
probability density functions $f_i(x_i)$.
In addition, if the random variables $X_i$ have discrete distributions then the results of this paper are still valid with
the integrals replaced by sums and the probability density functions replaced by the probability mass functions (see Hayter~(2014), for example).

General discussions of stochastic control can be found in Wendell \& Rishel~(1975) and 
{\O}ksendal~(2014), for example.
Moreover, in finance the problem of option pricing is also considered a stochastic control problem.  
Fusai \&  Meucci~(2008) discuss  pricing discretely monitored Asian options, 
and recursive integration techniques in pricing barrier options have been discussed in
Aitsahlia \& Lai~(1997), 
Sullivan~(2000), 
Andricopoulos et al.~(2003) 
and
Fusai \& Recchioni~(2007). 

The results obtained in this paper can also be used to calculate conditional probabilistic expressions and moments for
the stochastic behavior of these partial sum processes.
For example, probabilities for the independent random variables $X_i$ conditioned on an event $A$ can also  be tractable since
\begin{eqnarray}
P( (X_1, \ldots ,X_n) \in C | (X_1, \ldots ,X_n) \in A) = \frac{P( (X_1, \ldots ,X_n) \in C \cap A )}{P( (X_1, \ldots ,X_n) \in A)}
 \label{5}
\end{eqnarray}
where the numerator is tractable for certain sets $C$.
In particular, if the set $C \subseteq \Re^n$ can also be defined in terms of the partial sums as
\[
(X_1+ \ldots + X_i, X_{i+1},X_{i+2}) \in C_{i} \subseteq \Re^3
\]
for $1 \leq i \leq n-2$, then both the numerator and denominator of equation 
(\ref{5}) are of the form of equation 
(\ref{4})
with the sets $I_i$ equal to the sets $A_i \cap C_i$ or $A_i$ and the functions $h_i(x_i)$ equal to the 
probability density functions $f_i(x_i)$.

Furthermore, it can be noted that the moments and covariances of the 
independent random variables $X_i$ conditioned on an event $A$ are also  tractable since
\[
E [  X_1^{r_1} \ldots X_n^{r_n} \mid (X_1, \ldots ,X_n) \in A] = \frac{D}{P( (X_1, \ldots ,X_n) \in A)}
\]
where $D$ 
is of the form of equation 
(\ref{4})  with the sets $I_i$ equal to the sets $A_i$ and the functions $h_i(x_i)$ equal to  $x_i^{r_i}f_i(x_i)$.

The layout of this paper is as follows.
In section 2 it is shown how the integral in equation (\ref{4})
can be evaluated recursively as a series of $2$-dimensional integral calculations.
Recursive formulas are given for the general case, and also for the special case of equation (\ref{2}).
In section 3 a discussion is provided of the implementation details of the methodology.
The adoption of Fast Fourier Transforms  is illustrated as a way to improve the computational  efficiency of the methodology, 
and an error analysis of the numerical integrations is provided.
Some  illustrations of the implementation of the methodology are provided in
section 4, with examples in the fields of reliability, product quality assessment, and stochastic control.
Finally, a summary is provided in section 5.

\section{Recursive Integration Methodology}

In this section the recursive integration of the integral in equation (\ref{4}) is discussed.
First a change of variables is used to put the expression into a more convenient form, and then the general recursive formulas are provided.
The special case of equation (\ref{2}) is then considered separately.
It should be remembered that if the random variables $X_i$ have discrete distributions, then the results of this section are still applicable with
 the integrals replaced by sums, and the probability density functions replaced by the probability mass functions (see Hayter (2014), for example).

\subsection{Change of Variables}

If the change of variables $y_i = x_1 + \ldots + x_i$, $1 \leq i \leq n$, is employed, then 
equation (\ref{4}) becomes
\begin{equation}\label{6}
\idotsint\limits_{(y_1,\ldots,y_n) \in\Psi} h_1(y_1)h_2(y_2-y_1)\cdots h_n(y_n-y_{n-1})~
dy_1\cdots dy_n
\end{equation}
where the set 
$\Psi \subseteq \Re^n$
is defined by the conditions
\[
(y_i, y_{i+1},y_{i+2}) \in J_{i} \subseteq \Re^3
\]
for $1 \leq i \leq n-2$,
and where the set $J_{i}$ is derived from the set $I_{i}$
through the relationship
\[
(x_1+ \ldots + x_i, x_{i+1},x_{i+2}) \in I_{i} \Leftrightarrow (y_i, y_{i+1},y_{i+2}) \in J_{i} .
\]

Notice that in equation (\ref{6}) the integrand is the product of terms that only involve two adjacent $y_i$, while the integration region
is defined by conditions on only three adjacent $y_i$.
Consequently,  equation (\ref{6}) is of the form given in section 1 of Hayter (2006) with $d=2$,
which implies that it can be evaluated recursively by a series of 2-dimensional integral calculations.
Specific formulas for this recursive integration are now provided.

\subsection{General Recursive Formulas}\label{subsec:formula}

Let 
\[J_k(\cdot,u,v) = \{x\in\Re: (x,u,v)\in J_k\}. \]
We assume that $J_k(\cdot,u,v)$ can always be represented by a finite union of disjoint closed intervals, so that  
\begin{equation}\label{eqn:uinterval}
J_k(\cdot,u,v) = \cup_i [a_{k,i}(u,v), b_{k,i}(u,v)].
\end{equation}
In addition, let
\[J_k^{12} = \{(x,y)\in\Re^2: \exists z\in\Re \ni (x,y,z) \in J_k\},\]
and
\[J_k^{23} = \{(y,z)\in\Re^2: \exists x\in\Re \ni (x,y,z) \in J_k\}.\]
To compute equation (\ref{6}),  at each $(u,v)\in J_2^{12}$ first evaluate
\begin{equation}\label{8}
G_1(a,u) = \int_{-\infty}^a h_1(x)h_2(u-x)\,dx,
\end{equation}
for all $a\in \cup_v\cup_i[a_{1,i}(u,v), b_{1,i}(u,v)]$, and then compute
\begin{equation}\label{9}
g_1(u,v) = \int_{J_1(\cdot,u,v)} h_1(x)h_2(u-x)\,dx = \sum_{i} [G_1(b_{1,i}(u,v),u) - G_1(a_{1,i}(u,v),u)].
\end{equation}

Next, for $k=2,\ldots,n-3$, at each $(u,v)\in J_{k+1}^{12}$, and for $k=n-2$ at each $(u,v)\in J_k^{23}$, 
evaluate
\begin{equation}\label{10}
G_k(a,u) = \int_{-\infty}^a g_{k-1}(x,u)h_{k+1}(u-x)\,dx,
\end{equation}
 for all $a\in\cup_v\cup_i[a_{k,i}(u,v), b_{k,i}(u,v)]$, and letting $g_k(u,v)=0$ for $(u,v)\in J_k^{23}$ but $(u,v)\notin J_{k+1}^{12}, k = 2,\ldots,n-3$.
Then compute
\begin{equation}\label{11}
g_k(u,v) = \int_{J_k(\cdot,u,v)} g_{k-1}(x,u)h_{k+1}(u-x)\,dx = \sum_{i} [G_k(b_{k,i}(u,v),u) - G_k(a_{k,i}(u,v),u)].
\end{equation}
Finally, the evaluation of equation (\ref{6}), and hence of equation (\ref{4}), is obtained as
\begin{equation}\label{12}
P((X_1, \ldots ,X_n) \in A ) = \iint\limits_{J_{n-2}^{23}} g_{n-2}(u,v)h_n(v-u)\,du\,dv.
\end{equation}
Notice that the steps in this evaluation each have the computational intensity of a two-dimensional numerical integration.

\subsection{Recursive Formulas for the Sum of Independent Random Variables}

Now consider the special case where the $X_1,\ldots,X_n$ are independent random variables with 
probability density functions $f_1,\ldots,f_n$, respectively.  Recursive formulas are now provided for 
evaluating some probabilistic properties of $T = X_1+\cdots+X_n$. 
First, notice that it follows from equation (\ref{6}) that
\begin{equation}\label{eqn:sumint}
P(T\leq\tau) = \idotsint\limits_{y_n\leq\tau} f_1(y_1)f_2(y_2-y_1)\cdots f_n(y_n-y_{n-1})
\,dy_1\cdots dy_n .
\end{equation}
This expression can be computed simply by first evaluating  
\[g_1(u) = \int_{-\infty}^\infty f_1(x)f_2(u-x)\,dx\]
at each $u\in\Re$.
Then, sequentially, for $k=2,\ldots,n-1$, evaluate
\[g_k(u) = \int_{-\infty}^\infty g_{k-1}(x)f_{k+1}(u-x)\,dx\]
at each $u\in\Re$.
Again, notice that the steps in this evaluation each have the computational intensity of a two-dimensional numerical integration.
Finally, the required expression is obtained as
\[P(T\leq \tau) = \int_{-\infty}^\tau g_{n-1}(x)\,dx.\]

To compute the expectation of $w(X_1)$ conditional on $T\leq \tau$, or $T\geq\tau$,
first evaluate 
\[g_1^1(u) = \int_{-\infty}^\infty w(x) f_1(x)f_2(u-x)\,dx.\]
Then, sequentially, for $k=2,\ldots,n-1$, evaluate
\[g_k^1(u) = \int_{-\infty}^\infty g_{k-1}^1(x)f_{k+1}(u-x)\,dx\]
for each $u\in\Re$.
The expectation of $w(X_1)$ conditional on $T\leq \tau$ can then be obtained as
\[E[w(X_1)\mid T\leq \tau] = \frac{\int_{-\infty}^\tau g_{n-1}^1(x)\,dx}{P(T \leq \tau)} , \] 
while the expectation of $X_1$ conditional on $T\geq \tau$ can be obtained as
\[E[w(X_1)\mid T\geq \tau] = \frac{\int_{\tau}^\infty g_{n-1}^1(x)\,dx}{P(T \geq \tau)}.\]
Notice that expectations for $w(X_i)$ can be obtained from these expressions by reordering the indices of the $X_i$.

To compute the expectation of $w_1(X_1)w_2(X_2)$ conditional on $T\leq \tau$, or $T\geq\tau$, first evaluate 
\[g_1^2(u) = \int_{-\infty}^\infty w_1(x)w_2(u-x) f_1(x)f_2(u-x)\,dx.\]
Then, sequentially, for $k=2,\ldots,n-1$, evaluate
\[g_k^2(u) = \int_{-\infty}^\infty g_{k-1}^2(x)f_{k+1}(u-x)\,dx\]
for each $u\in\Re$.
The expectation of $w_1(X_1)w_2(X_2)$ conditional on $T\leq \tau$ can then be obtained as
\[E[w_1(X_1)w_2(X_2)\mid T\leq \tau] = 
\frac{\int_{-\infty}^\tau g_{n-1}^2(x)\,dx}{P(T \leq \tau)},\]
while the expectation of $w_1(X_1)w_2(X_2)$ conditional on $T\geq \tau$ can be obtained as
\[E[w_1(X_1)w_2(X_2)\mid T\geq \tau] = 
\frac{\int_{\tau}^\infty g_{n-1}^2(x)\,dx}{P(T \geq \tau)}.\]
Again,  expectations for  $w_i(X_i)w_i(X_j)$ can be obtained from these expressions by reordering the indices of the $X_i$.

Finally, notice that the expectation of $T$ conditional on either $T\leq \tau$ or $T\geq \tau$ is 
\[E[T\mid T\leq\tau] = \sum_{i=1}^n E[X_i\mid T\leq\tau]\quad
\text{and}\quad E[T\mid T\geq\tau] = \sum_{i=1}^n E[X_i\mid T\geq\tau], \]
which becomes
\[E[T\mid T\leq\tau] = nE[X_1\mid T\leq\tau]\quad
\text{and}\quad E[T\mid T\geq\tau] = nE[X_1\mid T\geq\tau]\]
when the $X_i$ are identically distributed.

\section{Implementation details}\label{sec:implement}

In this section
a discussion is provided of the implementation details of the methodology.
The adoption of Fast Fourier Transforms  (see Carverhill \& Clewlow (1990), for example) is illustrated as a way to improve the computational  efficiency of the methodology, 
and an error analysis of the numerical integrations is provided.

With $n$ variables $X_i$ a direct implementation of the methodology requires a calculation with a computational intensity that is equivalent to a 
sequence of $n$ two-dimensional numerical integrations.  
This is already  efficient considering that the original problem is ostensibly an $n$-dimensional numerical integration. 
However, in the case when the limits of integration $\cup_i \{a_{k,i}(u,v), b_{k,i}(u,v)\}$ in section 2.2 are invariant 
over pairs $(u,v)$,the computation can be accelerated even further using a 
Fast Fourier Transform  convolution.

\subsection{Fast Fourier Transform Convolution}
It can be observed that the recursive integration formulas given in section 2 involve the convolution of two functions.
Consequently, in some cases the
speed of the computation can be increased with a Fast Fourier Transform technique (FFT).  As the well-known Convolution Theorem states (see, for example, Smith (2007)), 
a convolution with respect to the variable in the original domain is equivalent 
to multiplication with respect to the variable in the transformed domain.

More formally, letting $F$ denote the Discrete Fourier Transform (DFT) and $F^{-1}$ its inverse,  convolutions between two functions $f$ and $g$ can be computed as
\[f {*} g = F^{-1} (F(f) \cdot F(g)).\]
The functions are decomposed into the transformed domain using the DFT, multiplied in the transformed domain, and then transformed back into the original domain using the inverse DFT.

Notice that the DFT and its inverse can be calculated by the FFT algorithm. 
Using a grid size of $N$, 
the overall computational intensity of conducting the convolution in this way using the FFT is $O(N \log N)$ (see, for example, Smith (2007)), 
which is lower than the computational intensity $O(N^2)$ obtained with the direct computation of the convolution in the time domain. The comparative accuracies and efficiencies of the two methods are now demonstrated.

\subsection{Accuracy and Efficiency}

In order to illustrate  and compare the accuracies and efficiencies of the implementations of the recursive integration formula introduced in section~\ref{subsec:formula}, the formula is applied to the calculation of the cumulative distribution function, the conditional cumulative distribution function, and the conditional expectation of the sum of 10 independent identically distributed exponential random variables with parameter $\lambda$ = 1. In this  case  the sum of these random variables has a known gamma distribution, so that the exact values of the calculated quantities are known.

The formulas in section 2.2 are implemented with a truncation of the support at 30.
Table~\ref{tab:001} shows the computed values and the errors of the required quantities, together with their computational times, obtained from implementations with the direct convolution and the FFT convolution. Different grid sizes are used, and both methods are implemented with SciPy's Python library (see Jones et al. (2001)) with an Intel Core i5 CPU.

\begin{table}[h!]
\caption{Comparisons of the implementation methods for the methodology for  a sum of 10 independent identically distributed exponential random variables with parameter $\lambda = 1$.}\label{tab:001}
\begin{center}
\begin{footnotesize}
\begin{tabular}{| r | l | c | c | c | c | c | c |}
\hline
 & & \multicolumn{3}{c|}{Direct Convolution} & \multicolumn{3}{c|}{FFT Convolution} \\
 \cline{3-8}
Grid size &   & Value & Error & Time (sec) & Value & Error & Time (sec) \\
\hline
0.01 & $P(T \geq 12)$ & 0.24275 & 0.00036 & 0.030 & 0.24275 & 0.00036 & 0.017 \\
\cline{2-8}
 & $P(T \geq 12\mid T\geq 10)$ & 0.52945 & 0.00013 & 0.028 & 0.52945 & 0.00013 & 0.017 \\
\cline{2-8}
 & $E[X_1\mid T\geq 10]$ & 1.27289 & 0.00032 & 0.055 & 1.27289 & 0.00032 & 0.033 \\
\hline
0.001 & $P(T \geq 12)$ & 0.24239 & 0.00000 & 2.376 & 0.24239 & 0.00000 & 0.098 \\
\cline{2-8}
 & $P(T \geq 12\mid T\geq 10)$ & 0.52932 & 0.00000 & 2.342 & 0.52932 & 0.00000 & 0.092 \\
\cline{2-8}
 & $E[X_1\mid T\geq 10]$ & 1.27320 & 0.00001 & 4.719 & 1.27320 & 0.00001 & 0.185 \\
\hline
0.0001 & $P(T \geq 12)$ & 0.24240 & 0.00000 & 388.769 & 0.24240 & 0.00000 & 1.235 \\
\cline{2-8}
 & $P(T \geq 12\mid T\geq 10)$ & 0.52932 & 0.00000 & 408.467 & 0.52932 & 0.00000 & 1.191 \\
\cline{2-8}
 & $E[X_1\mid T\geq 10]$ & 1.27320 & 0.00001 & 838.920 & 1.27320 & 0.00001 & 2.285 \\
\hline
\end{tabular}
\end{footnotesize}
\end{center}
\end{table}

It can be seen from Table 1 that the two implementations have similar errors, but  the implementation with the FFT convolution is significantly faster.
Consequently, it is useful to apply the FFT technique to the evaluation of the recursive formulas given in section 2.2 when 
the limits of integration $\cup_i \{a_{k,i}(u,v), b_{k,i}(u,v)\}$  are invariant 
over pairs $(u,v)$.

\section{Examples and Illustrations}

In this section the methodology presented in this paper is illustrated
through applications to  problems in the fields of reliability, product quality assessment, and stochastic control that require probability calculations for partial sums of independent random variables.
The first example concerns a reliability problem  where failed components are successively replaced with new components,
while the second example concerns a product quality assessment problem where batches are evaluated based on a measurement of
the sum of their individual items.
Finally, the third problem concerns discrete time stochastic control.

\subsection{Reliability Example}

Suppose that a machine contains $n$ ``identical'' components which are deployed successively.
Thus, the first component is deployed until it fails, whereupon the second component is deployed, and so on.
The machine operates until the $n$th component has failed.
Furthermore, suppose that an observer can tell whether or not the machine is operating, but not how many components have failed if the machine is still operating.

If the component lifetimes are taken to be independent with specified distributions, then the methodology presented in this paper can be used to
investigate the probabilistic properties of the lifetime of the machine.
Some illustrative calculations are provided when the component lifetimes are taken to be independent identically distributed Weibull
distributions.
Without the methodologies presented here, calculations on the sum of Weibull distributions are generally intractable
and would usually be assessed with simulations.

The following are examples  of the kinds of probability calculations that can be
performed using the recursive integration methodology presented in this paper.
If the component lifetimes are $X_i$ with distributions $f_i(x_i)$, so that the machine lifetime is $T = X_1 + \ldots + X_n$, then
an obvious quantity of interest is the machine survival function
\[
P( T \geq t).
\]
If the machine is observed at time $\tau$, then if the machine is still operating the conditional survival function is
\[
P( T \geq t\mid T \geq \tau) = \frac{P( T \geq t)}{P( T \geq \tau)}.
\]
If the machine has failed at time $\tau$ then the conditional survival function is
\[
P( T \geq t\mid T \leq \tau) = \frac{P( t \leq T \leq \tau)}{P( T \leq \tau)}.
\]

The expected failure time of the machine is simply $n$ times the individual expected component failure time, but if the machine is observed to be still operating at time $\tau$, then the conditional expected failure time is
\[
\int_{t=\tau}^{\infty} t f(t \mid t \geq \tau) dt = \frac{G}{P( T \geq \tau)}
\]
where $f(t | t \geq \tau)$ is the conditional distribution of the failure time and
\[
G = \int
  \!\!\!
\begin{array}{c c c}
 \\
 \ldots
 \\
x_1 + \ldots + x_n \geq \tau
\end{array}
\!\!
 \int
(x_1 + \ldots + x_n) f_1(x_1) \ldots f_n(x_n) ~ dx_1 \ldots dx_n .
\]
This can be evaluated as the sum of $n$ separate integrals which are identical if the component lifetimes are identically distributed.
The variance of the conditional  failure time can be obtained by having $t^2$ in place of $t$ in the integrand,
so that $G$ can be found from terms with $x_i^2$ and $x_i x_j$ in the integrand.

Finally, if the machine is observed to be still operating at time $\tau$, then the distribution of the number of failed components at time $\tau$
can be obtained, for $1 \leq i \leq n-1$, as
\[
P( \mbox{no more than $i-1$ components have failed by time $\tau$}) =
\]
\[
P( X_1 + \ldots + X_i \geq \tau \mid T \geq \tau) = \frac{P( X_1 + \ldots + X_i \geq \tau)}{P(  T \geq \tau)}.
\]

Table~\ref{tab:01}  shows the computed results (with computation times using the Fast Fourier Transform technique) of these probabilities  when 
$X_1,\ldots,X_{10}$ are independent, identically distributed Weibull random variables with shape parameter equal to 2 and scale parameter equal to 1.
These random variables have an expectation of 0.886 and a standard deviation of 0.463.

\begin{table}[h!]
\caption{Computed results and computation times for reliability example.}\label{tab:01}
\begin{center}
\begin{tabular}{| l | c | c |}
\hline
 & Computed value & Computational time (sec)  \\
\hline
$P(T \geq 8)$ & $0.7139490$ & $1.876$ \\
$P(T \geq 10)$ & $0.2154629$ & $2.040$ \\
$P(T \geq 12)$ & $0.0206421$ & $1.926 $ \\
$P(T \geq 12\mid T\geq 10)$ & $0.0958036$ & $1.701$ \\
$P(8 \leq T < 10\mid T\leq 10)$ & $0.6353888$ & $1.617$ \\
$P(X_1+\cdots+X_7\geq 10\mid T\geq10)$ & $0.0104016$ & $1.776$ \\
$E[T]\textsuperscript{\textdagger}$ & $8.8627912$ & $1.797$ \\
$E[T\mid T\geq 10]$ & $12.3020396$ & $3.664$ \\
\hline
\multicolumn{2}{l}{\textsuperscript{\textdagger}\footnotesize{$E[T]$ the exact value is equal to $10\Gamma(1.5)$}}
\end{tabular}
\end{center}
\end{table}

\subsection{Product Quality Example}

Consider a product quality assessment problem where a measureable property of an item is satisfactory if it is no smaller than a specified level $c$.
Let $X_i$, $1 \leq i \leq n$, represent the values of these properties for a batch of $n$ items,
and suppose that they can be modelled as being independent with an identical probability density function $f(x)$.

Suppose that instead of the costly approach of testing each item in the batch, it is possible and simple to obtain information about the sum $T = X_1 + \ldots + X_n$.
This is the case, say, if the weight of the item is of interest or the radiation emitted from the item.
It is useful to be able to make probability statements about the number of satisfactory items in the batch based upon the information obtained about $T$.
In practice, the exact value of $T$ may be observed, or a lower or an upper bound may be obtained.

If the  exact value of $T$ is observed then
\[
P(\mbox{exactly $i$ items are satisfactory}\mid T) = 
\]
\[
{n\choose i} P(X_1 \geq c , \ldots , X_i \geq c , X_{i+1} < c , \ldots , X_n < c | T) = {n\choose i} \frac{H_1}{H_2}
\]
where
\[
H_2 = \int
  \!\!\!
\begin{array}{c c c}
 \\
 \ldots
 \\
x_1 + \ldots + x_n = T
\end{array}
\!\!
 \int
f(x_1) \ldots f(x_n) ~ dx_1 \ldots dx_n
\]
and
\vspace{-5mm}
\[
H_1 = \int
  \!\!\!
\begin{array}{c c c}
\\
~~
 \\
~~
 \\
 \ldots
 \\
x_1 + \ldots + x_n = T
\\
x_1 \geq c , \ldots , x_i \geq c 
\\
x_{i+1} < c , \ldots ,x_n < c
\end{array}
\!\!
 \int
f(x_1) \ldots f(x_n) ~ dx_1 \ldots dx_n .
\]

As an illustration, some calculations are shown when $n=10$, $c=1$ and $f(x)$ is taken to be a Laplace (double exponential) distribution with 
parameter $\lambda=1$, so that
\[f(x) = \frac{1}{2}e^{-|x|}, x\in\Re.\]
Table~\ref{tab:02} shows the computed values  of $P(\mbox{exactly $i$ items are satisfactory}\mid T)$ 
at different  $i=0,1,\ldots,10$ and $T=0, 5, 10, 15, 20$.
The computational time of each entry using the Fast Fourier Transform technique was about 0.3 seconds.

If the bounds $T \leq t$ or $T \geq t$ are observed rather than the exact value of $T$,
then the expressions for $H_1$ and $H_2$ can be modified so that the integration regions depend on the conditions 
$x_1 + \ldots + x_n \leq t$ or $x_1 + \ldots + x_n \geq t$.
In either case $H_1$ and $H_2$ can be again be evaluated using the recursive integration methodologies presented in this paper.

\begin{table}[h!]
\caption{The probability of exactly $i$ items having a satisfactory weight (weight  greater than $c=1$) given an observed total weight $T$ of 
$n=10$ items.  The items are assumed to have independent and identically distributed weights with 
a Laplace distribution with parameter $\lambda=1$.  The computational time of each entry using the Fast Fourier Transform technique was about 0.7 seconds.
}\label{tab:02}
\begin{center}
\begin{tabular}{| c | r | r | r | r | r |}
\hline
 & \multicolumn{5}{c|}{$T$} \\
\cline{2-6}
 i & \multicolumn{1}{c|}{0} & \multicolumn{1}{c|}{5} & \multicolumn{1}{c|}{10} & \multicolumn{1}{c|}{15} & \multicolumn{1}{c|}{20}\\
 \hline
0&0.0774&0.0004&0.0000&0.0000&0.0000\\
1&0.3629&0.0518&0.0024&0.0002&0.0000\\
2&0.3896&0.2960&0.0477&0.0076&0.0016\\
3&0.1461&0.4176&0.2315&0.0688&0.0213\\
4&0.0225&0.1971&0.3905&0.2374&0.1135\\
5&0.0015&0.0347&0.2560&0.3568&0.2771\\
6&0.0000&0.0023&0.0656&0.2443&0.3310\\
7&0.0000&0.0001&0.0061&0.0751&0.1948\\
8&0.0000&0.0000&0.0002&0.0094&0.0542\\
9&0.0000&0.0000&0.0000&0.0004&0.0063\\
10&0.0000&0.0000&0.0000&0.0000&0.0002\\
\hline
\end{tabular}
\end{center}
\end{table}

\subsection{Discrete Time Stochastic Control Example.}

This section illustrates the application of the methodology developed in this paper to a discrete time stochastic control problem.  Let $X_i$, $i=1,\ldots,N$, be the performance measurement 
of a process at discrete times  $i$, where the $X_i$ are non-negative and assumed to be independent and identically distributed when the process is operating correctly.  
The objective is to dynamically track the partial means  of the $X_i$ over time, and to detect any increase in the mean of the $X_i$ by a certain decision rule.  

For $n=1,\ldots,N$, denote the partial means up to $n$ by
\[\bar{X}_n = \frac{\sum_{i=1}^n X_i}{n}.\]
Suppose that for each $n=3,\ldots,N$,  the process is stopped when both $X_{n}$ and $X_{n-1}$ 
are greater than $\bar{X}_{n-2}+c(\alpha, N)$, for a certain control limit $c(\alpha, N)$.  
If  the process is not stopped prior to $N$, then  the process is deemed to have been operating correctly throughout the time horizon $N$.  
For a specified distribution of the $X_i$, 
it is required to calculate the value of $c(\alpha, N)$ that  provides a probability of
$1-\alpha$ of not incorrectly stopping the process within the horizon $N$.

The control limit $c(\alpha, N)$ can be obtained by searching for the value of $c^*$ that is the solution to the equation
\begin{equation}\label{eqn:control}
P(X_{k+1} \leq \bar{X}_{k} + c^* \text{ or } X_{k+2}\leq \bar{X}_{k} + c^*,\text{ for } k=1,\ldots,N-2) = 1-\alpha.
\end{equation}
The event in equation (\ref{eqn:control}) is the event that the process is not terminated within the time horizon.  
This event is in  the form of equation (1), which  can be computed by the formula in equation (\ref{6}).  

In order to compute equation (\ref{6}), the $J_k(\cdot,u,v)$ in equation(\ref{eqn:uinterval}) have
 $u$ and $v$ as the transformed variables
\[Y_{k+1} = \sum_{i=1}^{k+1} X_i\quad\text{ and }\quad Y_{k+2} = \sum_{i=1}^{k+2} X_i.\]
Furthermore, given $Y_{k+1} = u$ and $Y_{k+2} = v$, 
the process is in control at time $k$ if
\[Y_k \geq \frac{k}{k+1}(u - c^*),\]
or
\[Y_k \geq k(v-u-c^*).\]
In addition, since the $X_i$ are non-negative random variables it follows that
\[Y_k \leq u,\quad\text{for }k=1,\ldots,N-1.\]
Therefore,
\[J_k(\cdot,u,v)=[a_k(u,v), b_k(u,v)],\] where
\[a_k(u,v) = \min\left\{\frac{k}{k+1}(u - c^*),\ k(v-u-c^*) \right\}\]
and
\[ b_k(u,v) = u.\]
Notice that in this case the Fast Fourier Transform technique cannot be used 
because the limits of the integrals $a_k(u,v)$ and $b_k(u,v)$ vary over $u$ and $v$.

To obtain the required control limit $c(\alpha, N)$, the probability in equation (\ref{eqn:control}) has to be computed at several values of 
$c^*$ in order to search for the solution.  Consequently, for a large time horizon $N$ it is essential that
an efficient computation methodology, as developed in this paper, is available in order to 
obtain $c(\alpha, N)$ in practice.

Table~\ref{tab:control} shows the control limit for different values of $\alpha$ and $N$, together with computational times using the recursive integration 
methodology developed in this paper, for the case where the $X_i$ are independent, identically distributed exponential random variables with scale parameter equal to 1.
\begin{table}[h!]
\caption{The control limit $c(\alpha, N)$ at $\alpha=0.05, 0.10$ and $N=8, 10, 12$.
}\label{tab:control}
\begin{center}
\begin{tabular}{| c | r | c | r | c | r | c |}
\hline
 & \multicolumn{6}{c|}{$N$} \\
\cline{2-7}
 $\alpha$ & \multicolumn{1}{c|}{8} & \multicolumn{1}{c|}{Time (sec)}  & \multicolumn{1}{c|}{10} &\multicolumn{1}{c|}{Time (sec)}  & \multicolumn{1}{c|}{12} & \multicolumn{1}{c|}{Time (sec)}  \\
 \hline
0.10 & 1.96 & 2300 & 2.28 & 2402 & 2.55 & 3980 \\
0.05 & 2.65 & 1964 & 3.08 & 2339 & 3.47 & 2804 \\
\hline
\end{tabular}
\end{center}
\end{table}

\section{Summary}

The tracking of the stochastic behavior of a partial sum process is an important problem.
 There are  many applications of partial sum processes, and in this paper examples have been provided   in the fields of reliability, product quality assessment, and stochastic control.

It has been shown how calculations of the probabilistic properties of such a partial sum process, which ostensibly  require an ability to compute high-dimensional multivariate probabilities, and so are consequently intractable in general, can in fact be solved as a sequence of 
two dimensional computations, with each computation being the convolution of two functions.

Finally, it has been shown how the Fast Fourier Transform technique can be utilized for the evaluation of these convolutions in some cases.
The results of this paper allow the efficient computation of the probabilistic properties of many important partial sum processes.

\newpage
\begin{center}
{\bf References}
\end{center}

\vspace{3mm} \noindent 
Aitsahlia, F. and Lai, T. L., 1997. ``Valuation of discrete barrier and hindsight
options,'' {\em The Journal of Financial Engineering}, 6 (2), 169-177.

\vspace{3mm} \noindent 
Andricopoulos, A.D., Widdicks, M., Duck, P.W. and Newton, D.P., 2003. 
``Universal option valuation using quadrature methods,'' 
{\em Journal of Financial Economics}, 67, 447-471.

\vspace{3mm} \noindent 
Belisle, C. J. P., Romeijn, H. E. and Smith, R.L., 1993. 
``Hit-and-run algorithm for generating multivariate distribution,'' 
{\em Mathematics of Operations Research}, 18, 255-266.

\vspace{3mm} \noindent 
Carverhill, A.P. and Clewlow, L.J., 1990. ``Flexible convolution'', {\em Risk}, 3, 25-
29.

\vspace{3mm} \noindent 
Dunnett, C. W. and Sobel, M., 1955. ``Approximations to the probability integral 
and certain percentage points of a multivariate analogue of {S}tudent's $t$-distribution,'' 
{\em Biometrika}, 42, 258-260.

\vspace{3mm} \noindent 
Fleming, W. H. and Rishel, R. W., 1975. `` Deterministic and Stochastic Optimal Control,'' Springer, New York.

\vspace{3mm} \noindent 
Fusai, G. and Meucci, A., 2008. ``Pricing discretely monitored Asian options under Levy processes,''  {\em Journal of Banking and Finance}, 32, 2076-2088.

\vspace{3mm} \noindent 
Fusai, G. and Recchioni, M.C., 2007. ``Analysis of quadrature methods for
pricing discrete barrier options,'' {\em Journal of Economic Dynamics and
Control}, 31 (3), 826-860.

\vspace{3mm} \noindent 
Hayter, A. J., 2006. ``Recursive integration methodologies with statistical applications,''
{\em Journal of Statistical Planning and Inference}, 136, 2284-2296.

\vspace{3mm} \noindent 
Hayter, A. J., 2014. ``Recursive formulas for multinomial probabilities with applications,'' 
{\em Computational Statistics}, 29 (5), 1207-1219.

\vspace{3mm} \noindent 
Jones E., Oliphant E., Peterson P., et al., 2001. ``SciPy: Open Source Scientific Tools for Python,''  http://www.scipy.org/.

\vspace{3mm} \noindent 
Kiatsupaibul, S., Hayter, A. J. and Wei, L., 2017. ``Rank constrained distribution and moment computations,'' {\em Computational Statistics and Data Analysis}, 105, 229-242.

\vspace{3mm} \noindent 
Kiatsupaibul, S. and Hayter, A. J., 2015. ``Recursive confidence band construction for an
unknown distribution function,'' {\em Biometrical Journal}, 57 (1), 39-51.

\vspace{3mm} \noindent 
Kiatsupaibul, S., Smith, R. L. and Zabinsky, Z. B., 2011. 
``An analysis of a variation of hit-and-run for uniform sampling from general region,'' 
{\em ACM Transactions on Modeling and Computer Simulation}, 21 (3), Article 16.

\vspace{3mm} \noindent 
Lov{\'a}sz, L., 1999. ``Hit-and-run mixes fast,'' {\em Mathematical Programming}, 86, 443-461.

\vspace{3mm} \noindent 
{\O}ksendal, B., 2014. ``Stochastic Differential Equations'', 6th Edition, Springer, Heidelberg.

\vspace{3mm} \noindent 
Smith, J. O., 2007.
``Mathematics of the Discrete Fourier Transform (DFT), with Audio Applications", Second Edition, W3K Publishing.

\vspace{3mm} \noindent 
Smith, R. L., 1984. ``Efficient Monte Carlo procedures for generating points uniformly 
distributed over bounded regions,'' {\em Operations Research}, 32, 1296-1308.

\vspace{3mm} \noindent 
Soong, W. C., and Hsu, J. C., 1997. ``Using complex integration to compute 
multivariate normal probabilities,'' 
{\em Journal of Computational and Graphical Statistics}, 6 (4), 397-415.

\vspace{3mm} \noindent 
Sullivan, M. A., 2000. ``Pricing discretely monitored barrier options'', 
{\em Journal of Computational Finance}, 3 (4), 35-52.

\end{document}